\documentclass[aps, prd, amsmath,amssymb,notitlepage, twocolumn, nofootinbib,superscriptaddress]{revtex4-1}
\usepackage{subfigure}
\usepackage{amssymb}
\usepackage{graphicx}
\usepackage{color}
\usepackage{amsmath, amsthm}
\usepackage[colorlinks=true,linkcolor=blue,citecolor=red,urlcolor=blue]{hyperref}
\begin{document}
\title{Optical Analogue of the Dynamical Casimir Effect in a Dispersion-Oscillating Fibre}
\author{Stefano Vezzoli}
\affiliation{School of Physics and Astronomy, University of Glasgow, Glasgow G12 8QQ, United Kingdom}
\affiliation{Institute of Photonics and Quantum Sciences, Heriot-Watt University, EH14 4AS Edinburgh, United Kingdom}

\author{Arnaud Mussot}
\affiliation{Universit\'e Lille, CNRS, UMR 8523-PhLAM-Physique des Lasers Atomes et Mol\'ecules, F-59000 Lille, France}

\author{Niclas Westerberg}
\affiliation{Institute of Photonics and Quantum Sciences, Heriot-Watt University, EH14 4AS Edinburgh, United Kingdom}
\affiliation{School of Physics and Astronomy, University of Glasgow, Glasgow G12 8QQ, United Kingdom}

\author{Alexandre Kudlinski}
\affiliation{Universit\'e Lille, CNRS, UMR 8523-PhLAM-Physique des Lasers Atomes et Mol\'ecules, F-59000 Lille, France}

\author{Hatef Dinparasti Saleh}
\affiliation{Institute of Photonics and Quantum Sciences, Heriot-Watt University, EH14 4AS Edinburgh, United Kingdom}

\author{Angus Prain}
\affiliation{School of Physics and Astronomy, University of Glasgow, Glasgow G12 8QQ, United Kingdom}
\affiliation{Institute of Photonics and Quantum Sciences, Heriot-Watt University, EH14 4AS Edinburgh, United Kingdom}

\author{Fabio Biancalana}
\affiliation{Institute of Photonics and Quantum Sciences, Heriot-Watt University, EH14 4AS Edinburgh, United Kingdom}

\author{Eric Lantz}
\affiliation{D\'epartement d’Optique P. M. Duffieux, Institut FEMTO-ST, UMR 6174 CNRS Universit\'e Bourgogne
Franche-Comt\'e, F-25030 Besan\c con, France
}

\author{Daniele Faccio}
\affiliation{School of Physics and Astronomy, University of Glasgow, Glasgow G12 8QQ, United Kingdom}
\email{daniele.faccio@glasgow.ac.uk}

\begin{abstract}
The dynamical Casimir effect is the generation of pairs of real particles or photons from the vacuum as a result of a non-adiabatic change of a system parameter or boundary condition. As opposed to standard parametric amplification where the modulation occurs both in space and in time, this fundamental process requires a pure modulation in time, which makes its detection particularly challenging at optical frequencies.
In this paper we experimentally demonstrate a realisation of the optical analogue of the mechanical dynamical Casimir effect in the near-infrared optical regime in a dispersion-oscillating photonic crystal fibre. The experiments are based on the equivalence of the spatial modulation of the fibre core diameter to a pure temporal modulation when this is considered in the co-moving frame of the travelling pump pulse. We provide evidence of optical dynamical Casimir effect by measuring quantum correlations between the spectrally resolved photon pairs. The non-classical nature of the measured light is supported by evidence of anti-bunching photon statistics.

\end{abstract}
\maketitle

{\bf{Introduction.}}
One of the most outstanding predictions of quantum field theory is that pairs of real particles can be generated from the vacuum as a result of a strong non-adiabatic change of a system parameter or boundary condition \cite{moore,davies}. 
This is referred to as the Dynamical Casimir Effect (DCE) \cite{moore,Yablonovitch1989,Schwinger1992,Dodonov,Nation2012} and is usually described as a process in which a cavity with periodically oscillating mirrors produces pairs of photons from the vacuum. For a sinusoidal modulation at frequency $\Omega$, the pair will be generated at frequency $\Omega/2$. 
Although the original DCE proposal was based on the mechanical movement of mirrors, an analogue effect can be achieved by simply imposing a time dependence on one of the system parameters. Indeed a recent work \cite{PRX2018} divides all DCE theoretical proposals into two groups: experiments where the pair generation is due to a mechanical movement of mirrors, named \textit{mechanical DCE}, and setups where the boundary conditions are changed without moving mirrors, but simply by modulating a system parameter, hence the label \textit{parametric DCE}. We prefer to refer to this second class of proposals as \textit{analogue} DCE, due to historical reasons, but the two terms are interchangeable.
An alternative to oscillating mirrors is an optical cavity either filled with or constituted by a medium of refractive index $n$: a periodic change of $n$ in time (and that is uniform in all spatial coordinates) is equivalent to a periodic change of the boundary conditions, as pointed out by Lambrecht \cite{Lambrecht} and Mendon\c{c}a \cite{Mendonca-book,mendonca}. Of particular relevance to this work, Lambrecht's proposal relies on the use of an optical nonlinearity to achieve the periodic modulation of $n=\sqrt{\varepsilon}$ where the dielectric permittivity is modulated by an oscillating electric field, $E$, through the second order susceptibility $\chi^{(2)}$: $\varepsilon = \varepsilon_0+\chi^{(2)}E$ \cite{Lambrecht}. The same physics can be realised also in a medium with a third order, $\chi^{(3)}$, nonlinearity \cite{carusotto}. The key observation is that for the case in which the $\chi^{(2)}$- or $\chi^{(3)}$-medium is shorter than the driving $E$-field wavelength, the medium will oscillate uniformly and the parametric amplification of photons from the vacuum state takes on a formal analogy with the mechanical DCE. \\
Yablonovitch \cite{Yablonovitch1989} noted the deep connection between DCE and parametric amplification, a connection which has been deeply analysed more recently by Nation et al. \cite{Nation2012}. It is thus worth to point out that the physics of DCE and parametric amplification, for instance spontaneous four wave mixing in a $\chi^{(3)}$ medium, carries more than a formal resemblance. 
The distinguishing physical feature between standard parametric amplification and DCE is that the former relies on a polarisation wave propagating in a long medium so the modulation occurs both in (longitudinal propagation) space and in time whereas the latter refers to a temporal variation of the medium that is uniform along the longitudinal propagation direction. This last condition can be obtained for example, as mentioned above, by ensuring that the medium is significantly thinner than the wavelength of the input $E$-field driving the polarisation wave \cite{Lambrecht,carusotto}. 
This subtle difference renders the experimental realisation of a DCE experiment particularly challenging as it would require large (to overcome the low amplification gain from sub-wavelength films) and fast (non-adiabatic) modulations of the refractive index. Recent experiments demonstrated analogue DCE photon pair emission at low frequency (5 GHz) in superconducting circuits \cite{Wilson2011} and in a Josephson metamaterial \cite{Lahteenmaki2012}, and DCE-like emission (albeit with no quantum signature in the emission) was reported for acoustic waves in a Bose-Einstein condensate \cite{Westbrook2012}. However, the generation of photon pairs by DCE in the optical region, e.g. by nonlinear optical process, has not been demonstrated to date.

In this paper we experimentally demonstrate a realisation of the DCE in the near-infrared optical regime mediated by the $\chi^{(3)}$ nonlinearity in a dispersion-oscillating photonic crystal fiber. Photonic crystal fibers with a periodic modulation of the group-velocity dispersion (GVD) are also called dispersion oscillating fibers (DOFs) and they are obtained by periodically modulating the core diameter during fabrication \cite{Droques2012,fabio1,fabio2,mussotmodulation2018}. We first describe the longitudinal spatial modulation of the fibre core diameter as a temporal modulation when this is considered in the co-moving frame of the travelling pump pulse. We then provide experimental evidence of optical DCE by measuring quantum correlations between the spectrally resolved photon pairs, expressed by the conventional coincidence-to-accidental ratio. Finally, in order to demonstrate the non-classical nature of the measured light, we provide evidence of anti-bunching photon statistics.\\ 
\begin{figure}
\centering
\includegraphics[width=8.5cm]{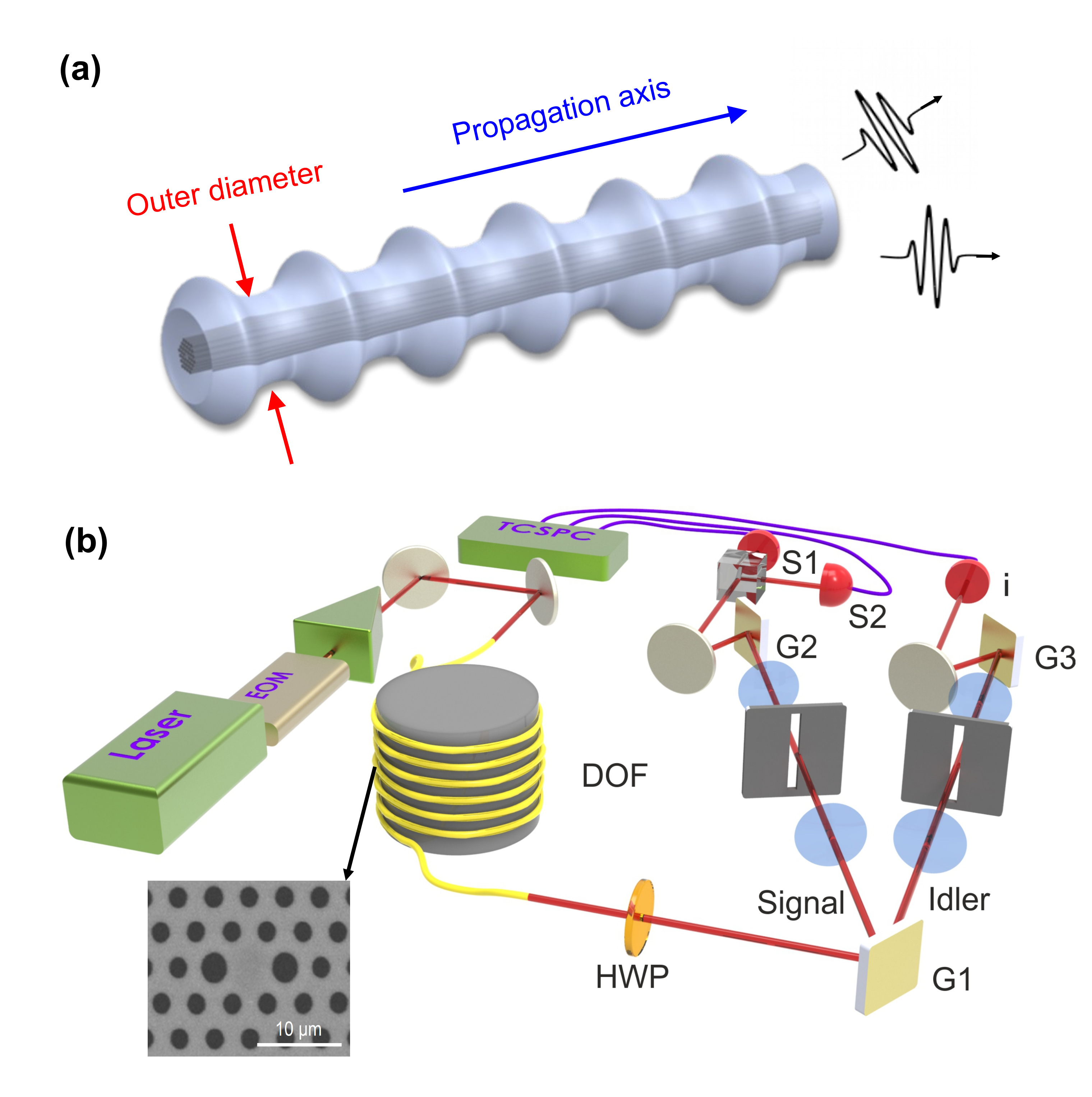}
\caption{(a) Concept of the Dynamical Casimir Effect in a dispersion oscillating fiber (DOF): a short pulse propagating though the fiber experiences a fast modulation of the group velocity dispersion (GVD). (b) Schematics of the experimental setup for quantum correlation measurements: signal and idler beams generated inside the DOF are separated, filtered from the pump by a 4-f grating system and detected by single photon avalanche detectors (SPAD) named s1,s2 and i. \label{fig:1}}
\end{figure}

{\bf{Theory.}}
Photon pair generation in fibers has a long history and is usually interpreted in terms of spontaneous four-wave mixing (SFWM). Parametric amplification of the vacuum fluctuations in a fiber was first realised by optically pumping with laser pulses whose carrier wavelength was chosen to be close to the zero dispersion wavelength of the fiber \cite{Li2004}. Phase matching and high efficiency photon pair generation was thus achieved by the balance between nonlinear phase contributions and the linear anomalous dispersion. Successively, the possibility to achieve phase matching was also shown in the normal dispersion regime \cite{Rarity2005} as a result of a negative fourth order dispersion term.\\
Parametric amplification (in the classical regime) in fibres with periodic spatial perturbations \cite{matera} and DOFs \cite{mussotmodulation2018,droquesb4} has also been observed and interpreted in terms of quasi-phase matching, in analogy to quasi-phase matched spontaneous parametric down-conversion in $\chi^{(2)}$ nonlinear crystals \cite{Boyd-book}.\\ 
{A schematic layout of the fibre geometry use in this work is shown in Fig.~\ref{fig:1}(a), where we consider the specific case in which the optical pump pulse is significantly shorter than the periodicity $\Lambda$ of the DOF oscillation.}
We then consider the evolution of the boundary conditions as perceived by such a short pulse in the reference frame of the pulse itself. The pulse will experience a uniform oscillation in {\emph{time}} of the surrounding medium parameters at a frequency $\Omega'$ that is proportional to the fibre longitudinal periodicity $K=2\pi/\Lambda$ (primed quantities refer to the frame comoving at the group velocity $v_g$ of the laser pulse). In this reference frame, the DCE predicts that photons will be generated at frequencies $|\omega'|=m{\Omega}'/2$, where the integer $m$ accounts for the possibility to have resonances also at frequencies that are multiples of the boundary modulation. If the medium has no optical dispersion, the phase velocity is equal to the group velocity $v=v_g$ and thus in the comoving frame the electric field does not oscillate in time. In this case, the only contribution to any time-variation in the comoving frame originates from the periodic fibre oscillation acting on the nonlinear refractive index $\Delta n\propto \chi^{(3)}|E|^2$. The presence of dispersion $v \neq v_g$ will lead to a slip of the pulse electric field $E$ underneath the pulse envelope, generating an additional temporal oscillation due to the oscillating electric field. This in turn creates an additional nonlinear polarisation term that is proportional to $\chi^{(3)}E^2$ and thus oscillates at twice the pulse comoving frequency $2\omega_0'$. In the case of our dispersive fibre therefore we have a modified DCE condition that must account for both temporally oscillating terms, i.e. $\omega'=m\Omega'/2+\omega_0'$.
In order to determine the emitted frequencies that will be observed in the laboratory frame, we Doppler shift this frequency condition from the comoving to the laboratory frame by
 using $\omega'=\gamma(\omega-v_gk)$ and $\Omega'=\gamma(\Omega-v_gK)=-\gamma v_gK$, where $\gamma=1/\sqrt{1-v_g^2/c^2}$ and $\Omega=0$ (in the lab frame the fibre does not oscillate in time). {The DCE condition therefore becomes:
\begin{equation}
\Delta\omega-v_gk=-mv_gK/2-v_gk_0
\end{equation}
where $\Delta\omega=\omega-\omega_0$. We now account for the fiber dispersion: 
\begin{equation}
k(\omega)\simeq k_0+\frac{\Delta\omega}{v_g}+\frac{1}{2}\beta_2\Delta\omega^2+\frac{1}{6}\beta_3\Delta\omega^3+\frac{1}{24}\beta_4\Delta\omega^4,
\end{equation}
where $\beta_n$ is the $n^{th}$ $\omega$-derivative of $k$ (averaged over the fiber modulation). Substituting this expansion into the condition for DCE emission, we find $\beta_2\Delta\omega^2+\frac{1}{3}\beta_3\Delta\omega^3+\frac{1}{12}\beta_4\Delta\omega^4=mK$. We must also account for energy conservation, so the positive and negative roots of this equation (frequency shifts $\Delta\omega$ of the generated photons with respect to the pump) must be equal in modulus, which gives (e.g. summing the equations for $+\Delta\omega$ and $-\Delta\omega$):
\begin{equation}
\beta_2\Delta\omega^2+\frac{1}{12}\beta_4\Delta\omega^4=mK
\label{DCE}
\end{equation} 
This expression predicts that the DCE photons will be observed in the lab frame in symmetric sidebands around the pump frequency} and provides a quantitative estimate of the exact spectral location of these photons. Interestingly, this formula, derived in the comoving frame as a DCE, is in perfect agreement with the result from a calculation based on the quasi-phase-matching condition for standard parametric amplification in the lab frame \cite{mussotmodulation2018,droquesb4} and therefore underlines once again the connection between the DCE and parametric oscillation.\\ 
\begin{figure}
\centering
\includegraphics[width=8.5cm]{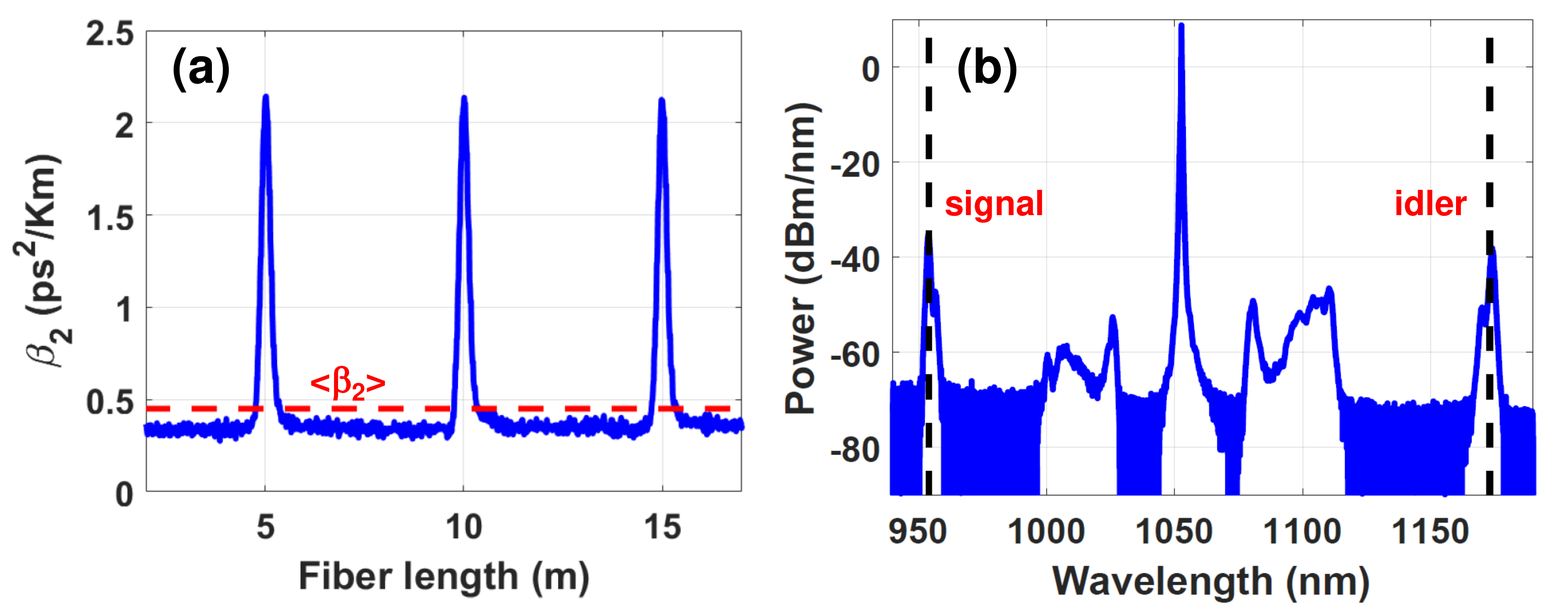}
\caption{(a) Zoom of the longitudinal evolution of the measured GVD with average value $<\beta_2>=0.45~ps^2/km$ for $\lambda_P=1052.44$ nm. The total length of the fiber is 80 $m$. (b) Optical spectrum measured at the output of the DOF (solid blue line) for high pump power $P_p=12 W$ and theoretical prediction from Eq.~\ref{DCE} (dashed black lines) for the third harmonic of the modulation frequency ($m=3$). \label{fig:2}}
\end{figure}

\begin{figure}
\centering
\includegraphics[width=8.5cm]{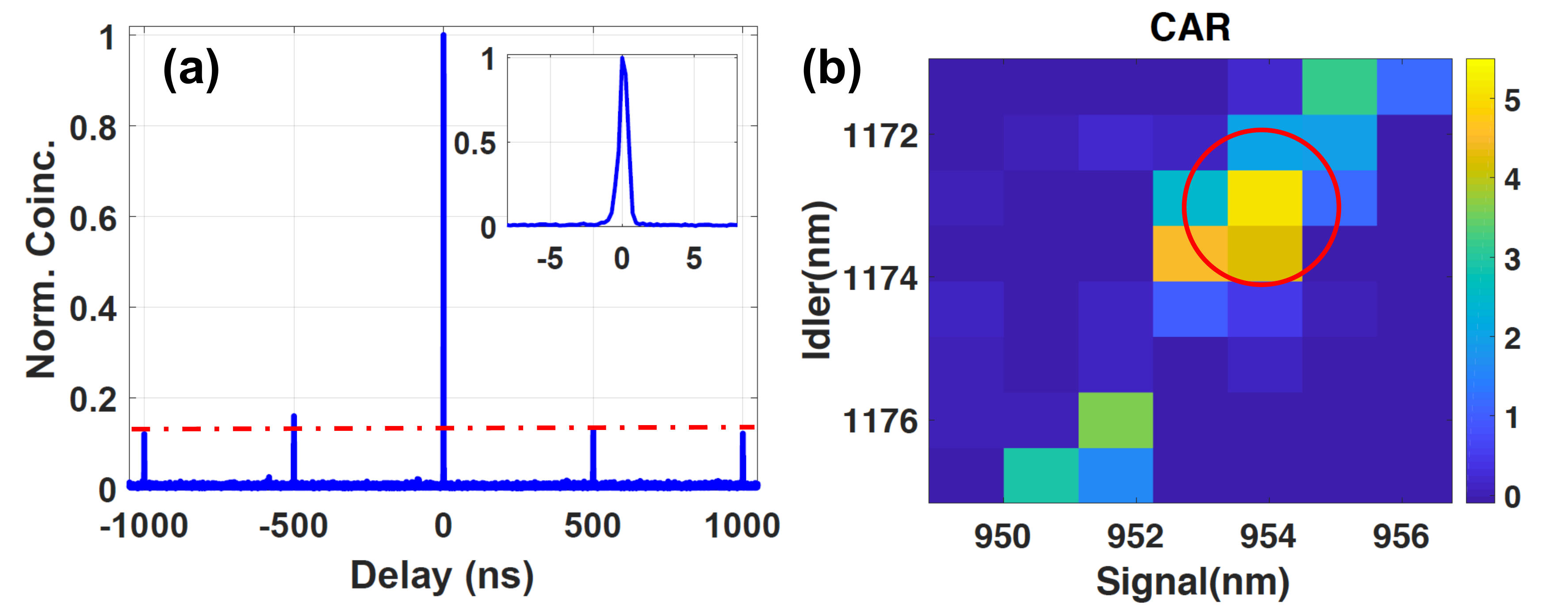}
\caption{(a) Example of CAR for $P_p=0.03$ W. The accidental counts are estimated from the coincidences between signal and idler at delay $\tau$ corresponding to multiples of the inverse of repetition rate. (b) 2D map of the CAR as a function of the signal and idler wavelengths for $P_p=0.03$ W and a coincidence time window $\Delta t=1.7$ ns. \label{fig:3}}
\end{figure}

\begin{figure}
\centering
\includegraphics[width=8.5cm]{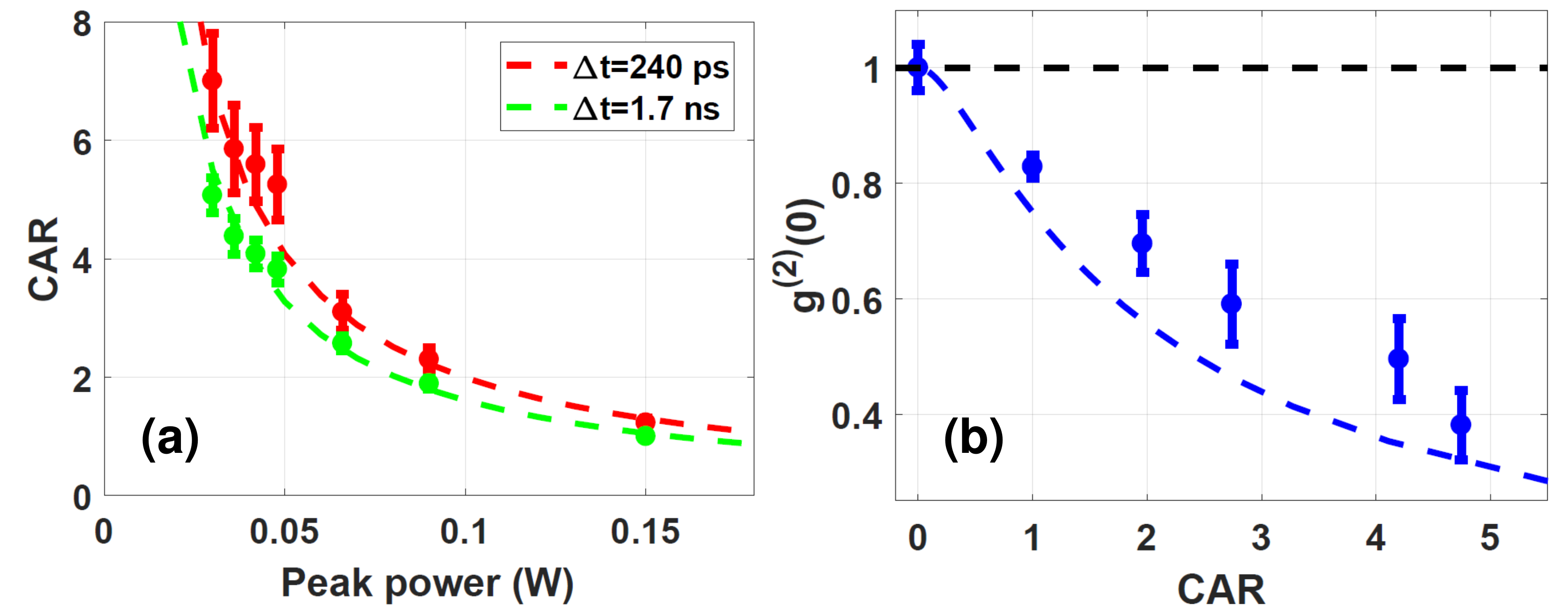}
\caption{ (a) CAR as a function of power for two different choices of the coincidence time window, $\Delta t=240$ ps and $\Delta t=1.7$ ns. The dashed lines are simulated for different ratios of Raman/DCE contributions on the idler channel. (b) Intensity auto-correlation function $g^{(2)}(0)$ at zero delay. The blue dashed line is a simulation assuming only single photon states $|1\rangle$ from DCE and Raman photons. \label{fig:4}}
\end{figure}

{\bf{Quantum emission measurements.}}
The GVD modulation of the photonic crystal fiber used in the experiments is illustrated in Fig.~\ref{fig:2}(a), whereas Fig.~\ref{fig:2}(b) shows a spectrum taken at high pump power, together with the prediction of the spectral sidebands from Eq.~\ref{DCE} for $m=3$ (dashed black lines at 954 $nm$ and 1173 $nm$, conventionally named signal and idler). More information about the fiber fabrication and classical characterisation are provided in the Methods section and in the Supplementary Information. 
Figure~\ref{fig:1}(b) shows a schematic view of the experimental setup used for quantum emission and correlation measurements. Diffraction gratings are used to filter out the pump power and to spectrally separate signal and idler beams. A spectral bandwidth of 1 $nm$  on both channels is selected in order to maximise the collection of DCE pairs and to minimise the residual contribution (noise) due to Raman scattering (see Methods for more details on the setup). The electronic signals generated by single photon detectors (SPADs) are time-stamped and correlations between signal and idler are measured by a time-to-digital converter (TDC) module. A histogram of coincidences versus the delay between the signal (s1 or s2) and idler (i) channels is shown in Fig.~\ref{fig:3}(a). The observed large peak of coincidence counts between signal and idler $N_{s,i}(0)$ at zero delay (i.e. within the same pump laser pulse) that is several times larger than the coincidence rates at different delays (i.e. between different laser pulses) unambiguously implies non-classical correlations between the signal and the idler beams \cite{Loudon}. The coincidence-to-accidental ratio (CAR) is defined as the ratio between the coincidences due to photon pairs and the accidental coincident counts. It can be estimated (see Supplementary Information for more details) as:
$$\frac{N_{s,i}(0)-N_{s,i}(\tau)}{N_{s,i}(\tau)}$$
where $N_{s,i}(0)$ is the coincidence peak at zero delay and $N_{s,i}(\tau)$ is the average of the non-zero-delay peaks. \\
In Fig.~\ref{fig:3}(b) we show the measured CAR as a function of the signal and idler wavelengths, obtained by scanning the signal and idler slits after the grating G1 with a step resolution of 1 nm. It is evident that the CAR remains large for wavelength-pairs that satisfy Eq.~\eqref{DCE} but otherwise quickly drops to zero. The best CAR (around 5) and the highest photon count rates are found for $\lambda_s=954$ $nm$ and $\lambda_i=1173$ $nm$, and this choice of wavelengths position will be used in the following analysis. The double-peak structure observed in Fig.~\ref{fig:3}(b) and similarly in Fig.~\ref{fig:2}(b) is ascribed to hopping of the pump laser between two modes of the laser cavity. \\
In Fig.~\ref{fig:4}(a) the CAR is measured for different pump peak powers between $0.03$ W and $0.15$ W and is seen to decrease with increasing power. This is due to the fact that accidental counts grow quadratically with the number of single-photon counts (originating from both DCE and Raman amplification), whereas the true coincidence counts grow only linearly. The estimated value of the CAR is dependent also on the time window, $\Delta t$, within which coincidences are counted and increases as we decrease $\Delta t$ [compare the green and red curves in Fig.~\ref{fig:4}(a)]. A very narrow time window of 240 ps allows us to collect most of the coincidences, while filtering out most of the background Raman and dark counts. \\
At very low powers on the idler channel most of the counts come from Raman scattering (Raman scattering emission occurs only at red-shifted wavelengths), whereas on the signal channel most of the single counts are due only to the detector dark counts. Therefore we use a model, described in details in Supplementary Information, in order to isolate the contribution of DCE pairs. We assume a quadratic dependence for the DCE photon-pair production process (2 photons from the pump are annihilated for each pair produced) and a linear dependence for the Raman process.
The dashed lines in Fig.~\ref{fig:4}(a) correspond to the resulting calculations based on the detected single photon rates, the estimated collection and detection efficiency and by using the ratio between Raman and DCE photons as a free parameter. From these calculations we estimate that about $2\times10^{-3}$ DCE pairs per pump pulse are generated in the fiber versus 0.18 Raman photons for $P_p=0.03$ W, or $0.05$ DCE pairs versus $0.9$ Raman photons at $P_p=0.15$ W.\\ 
We use these numbers to verify that the measured CAR is due to vacuum seeded photons and cannot be ascribed to seeding by the spontaneous Raman emission. This can be demonstrated by estimating the number of temporal modes contained in 1 nm of detected spectrum. From the Fourier transform of the 600 ps pump pulse we estimate a pump bandwidth of 3 GHz, to be compared with a detection bandwidth of about 300 GHz. Therefore we estimate roughly 100 temporal modes detected in 1 nm of spectrum. With 0.18 to 0.9 Raman photons per pulse, we therefore have between $1.8\times10^{-3}$ and $9\times10^{-3}$ Raman photons per temporal mode at the fiber output that are negligible when compared to the $1/2$-photon/mode from the vacuum, thus supporting the quantum vacuum-origin of the observed DCE photon coincidence counts.\\
%
Finally, we perform a heralded Hanbury-Brown Twiss experiment by using a beamsplitter on the signal path and measuring the coincidences at the two output ports, heralded by the idler photons. The second-order coherence at zero delay $g^{(2)}(0)$ is then evaluated as \cite{Bocquillon}:
$$ g^{(2)}(0)=\frac{N_{s1,s2,i}N_i}{N_{s1,i}N_{s2,i}}$$
where $N_i$ indicates the measured single count rate at the idler channel, $N_{x,y}$ the measured coincidence rates between the two beamsplitter ports $x=s1$ or $x=s2$ on the signal channel and the idler $y=i$, $N_{s1,s2,i}$ the triple coincidences between the three channels. $ g^{(2)}(0)<1$ is taken to be evidence of non-classicality \cite{Bocquillon}.\\
The results are shown in Fig.~\ref{fig:4}(b) for different values of CAR, corresponding to different pump powers. The case of CAR=0 is obtained by moving the idler slit in order to collect only Raman radiation on the idler and the corresponding $g^{(2)}(0)$ is found to be nearly equal to 1, as expected. 
The dashed blue line represents the calculated $g^{(2)}(0)$ in the case of only pure single photon states due to DCE pairs (the derivation is given in Supplementary Information). All experimental points lie slightly above the calculated curve, indicating a small contribution in the measurements from higher photon number states. The main result of Fig.~\ref{fig:4}(b) is that $g^{(2)}(0)$ clearly drops below 1 for CAR$>1$, thus providing a clear indication of non-classical emission.\\

{\bf{Conclusions.}} We have proposed a dispersion-oscillating photonic crystal fibre system for observing an optical realisation of the Dynamical Casimir Effect. Our proposal represents an optical analogue of the mechanical DCE in the sense that the mechanical motion of mirrors is replaced by a time modulation of a system parameter, the refractive index.
A pump laser pulse creates a refractive index variation through the nonlinear Kerr effect, which in the comoving frame experiences a temporal oscillation and, under the condition that the pulse is shorter than the fibre oscillation period, produces photon pairs according to the DCE condition.  The time oscillation of the fiber in the co-moving frame provides the mechanism to transfer energy from the pump beam to the vacuum states, in the same way as the modulation of the position of mirrors in a cavity allows the transfer of mechanical energy into photon pairs.
Experiments confirm the presence of correlated photon pairs emitted at the predicted wavelengths with non-classical statistics, supporting the origin from vacuum-seeded DCE. Therefore we propose that dispersion oscillating fibers could provide a suitable test-bed for studying the DCE physics, for instance by changing the shape of the fiber modulation. Moreover we showed that correlated pairs are produced with a narrow spectral bandwidth at frequencies which can be easily tuned by acting on the shape and period of the fiber modulation, thus providing a novel source of light for quantum applications. 

{\bf{Methods.}}\\
{\bf{Optical Fiber characterisation.}}\\
The fused silica photonic crystal fiber is fabricated by using the stack-and-draw technique \cite{russelPCF2003}. The outer diameter of the fiber, whose length is 80 $m$, is modified during the drawing process to obtain a Gaussian-shaped modulation of the GVD with a full width at half maximum of 0.5 m, repeated with a periodicity of 5 m as shown in Fig.~\ref{fig:2}(a). The average GVD value is $<\beta_2>=0.45$ ps$^2$/km (red dotted line in Fig.~\ref{fig:2}(a)) and the maximum excursion of GVD at the top of each spike is estimated to be about $\beta_2^{Max}=1.7$ ps$^2$/km. We pump the DOF with a fiber-amplified laser with a carrier wavelength $\lambda_P=1052.44$ nm and 2 MHz repetition rate. The pulse duration is 600 ps, equivalent to a length of 0.12 m (accounting for the refractive index $n$=1.45 of the fibre core) and is thus much shorter than the 5 m modulation periodicity.
The specific shape of the modulation is chosen in order to maximise the parametric gain at frequencies that lie far from the Raman gain spectrum. Experiments have also been carried out with a sinusoidal modulated fiber (corresponding to the simplest kind of modulation in DCE), and similar results to those described in this paper were found, although with slightly worse evidence of non-classical behaviour (lower CAR values). Fig. \ref{fig:2}(b) provides an example of the fiber output spectrum for a pump peak power $P_p=12$ W acquired with an optical spectrum analyser, showing parametric gain narrow peaks around 954 nm (signal) and 1173 nm (idler) and the broader contribution of Raman scattering mainly around 1100 nm. The dashed line corresponds to the prediction for DCE emission from Eq.\ref{DCE}, by using the experimental parameters measured for $\lambda_P=1052.44$ nm, $<\beta_2>=0.45$ ps$^2$/km and $<\beta_4>=-1.2\times 10^{-55}$ s$^4$/m (see Supplementary Information for more details).\\
{\bf{Experimental setup for quantum correlation measurements.}}
Fig.~\ref{fig:1}(b) shows the experimental setup used for quantum correlation measurements. We use a diffraction grating (G1) to filter out the pump spectrum and to also spectrally separate the photon pairs. Two additional gratings (G2 and G3) in a 4-f configuration are used as tunable passband filters in order to reduce the contribution of broadband residual Raman scattering. Slits are arranged in order to select 1 nm of spectrum in both the low (idler) and high frequency (signal) photon paths, which roughly corresponds to the DCE bandwidth, as illustrated in Fig.~\ref{fig:2}(b) and Fig.~\ref{fig:3}(b). The slit size (400 $\mu m$) was indeed verified to optimise the observed coincidence-to-accidental ratio (CAR). On the signal photon path we add a 50:50 beam splitter to perform anti-bunching measurements. Finally, photons are coupled into multi-mode fibers and sent to single-photon avalanche detectors (SPADs). We use two InGaAs gated SPADs (Aurea Technology) and one visible silicon SPAD (Excelitas). The outputs from the single photon detectors are sent to a time-to-digital converter (from IDQuantique), which registers the arrival times of the photons which are used to perform correlations. We operate the pump laser at low powers between $P_p=0.03$ W and $P_p=0.15$ W in order to ensure low gain and thus single photon-pair emission.\\

{\bf{Acknowledgements.}}
This work was partly supported by the Agence Nationale de la Recherche through the
High Energy All Fiber Systems (HEAFISY), the Labex Centre Europeen pour les Mathematiques, la
Physique et leurs Interactions (CEMPI) and Equipex Fibres optiques pour les hauts flux
(FLUX) through the Programme Investissements d'Avenir, by the Ministry of Higher
Education and Research, Hauts de France council and European Regional Development
Fund (ERDF) through the Contrat de Projets Etat-Region (CPER Photonics for Society,
P4S) and FEDER through the HEAFISY project.\\
D. F. acknowledges financial support from the European Research Council
under the European Union Seventh Framework Programme (FP/2007-2013)/ERC GA 306559 and EPSRC (UK, Grant No. EP/P006078/2). N. W. acknowledges support from EPSRC CM-CDT Grant No. EP/L015110/1.


\end{document}